\documentstyle[psfig,conf_iap,art10]{article}
\begin{document}
\heading{%
%
What can we learn from VIRMOS quasars?
%
} 
\par\medskip\noindent
\author{%
Guy Mathez, Evanthia Hatziminaoglou
}
\address{%
Laboratoire d'Astrophysique, Observatoire Midi-Pyr\'en\'ees, 14 Avenue E. Belin 31400 Toulouse, France \\
}

\begin{abstract}
Large, homogeneous quasar samples are necessary tools for the study of
QSO statistics, cosmological tests, large scale structure and AGN
evolution. These samples must be complete within well defined flux
limits at all redshifts. An observational strategy without previous
photometric selection of quasar candidates is described, 
based essentially on the VIRMOS Survey spectroscopy.
\end{abstract}
\section{Samples of quasars: why? }
Thanks to their high luminosities, quasi-stellar shape and 
strong emission lines, both photometry and spectroscopy of quasars are 
rather easy. The average redshift of quasar samples 
is still relatively high
in spite of the continuous increase of galaxy redshifts, so,
large and homogeneous quasar samples are useful for cosmological
purposes, in particular for the study of: i) the statistics of the 
AGN, which will help to progress towards a better understanding of the 
'unified' models, ii) large scale structure (clustering of QSOs and of 
QSO absorbers), and iii) for the global evolution of the AGN population 
and its links with normal galaxies. 
Recent observational data on the evolution of the density
of quasar light in the universe both from radio and optical quasars
show a reversing of the evolution around $z=3$ (\cite{S98}).
Since 1993, the corresponding theoretical understanding is in 
continuous progress, in terms of Eddington-limited growth of Super 
Massive Black Holes embedded in growing dark halos, 
(eg. \cite{HR}; \cite{H98}; \cite{SR}), undergoing successively 
turning-on and turning-off on short time-scales (\cite{HL})
through either viscous instabilities (\cite{SE}) or
tidal interactions (\cite{CV}). The latter authors
find 'negative' density evolution (DE) at $z>3$, corresponding to the
birth of galaxies, and 'positive' luminosity evolution (LE) at $z<3$,
corresponding to the progressive depletion of material to accrete.

\section{ Cosmological Tests with Quasars }
With a complete quasar sample, it is possible to perform
geometrical cosmological tests.
\cite{DXF94} compute a characteristic 
scale $\lambda_{QQ}\simeq 100 \; Mpc $, from the 
quasar-quasar correlation function, fix $\lambda_{QQ}=125 Mpc$ 
and constrain $( \Omega _m ,\Omega _{\Lambda})$
almost independently of the evolution. 
Eliminating anisotropies in the power spectrum of 
the 3D correlation function computed from redshift surveys 
of galaxies and quasars 
constraints on $\Lambda$ and $\Omega^{0.6}/b$
\cite{Bal96}, \cite{S96}. \cite{Bal97} analyse the
angular size-redshift relation of double-lobed FIRST quasars. 

Given a data set, any assumption on the
cosmological parameters leads to constraints on evolution, and 
conversely, as was done from quasar counts assuming Pure Luminosity
Evolution (PLE) \cite{SH} or Pure Density Evolution (PDE).

Assuming constant PLE parameter $k_L$ over 
the redshift range $[0.3,2.2]$, the V/Vmax Statistics applied to the AAT
sample (\cite{B90}) 
favors the couple of values $(\Omega=0.5\pm 0.3, \; 2 \sigma ; \; 
\Lambda=0.6\pm 0.4)$ (Fig. 1a) \cite{LVW96}. (quite
compatible with the recent results shown in this conference
by both SN teams)

Perhaps the most questionable hypothesis in LVW is that 
the luminosity evolution parameter
$k_L$ is constant over the redshift range $[0.3,2.2]$. 
Fortunately, it has been shown that either PLE or PDE 
with a roughly constant k is not too
bad an approximation, at least in the limited redshift range $0.7 \le z
\le 1.7$ (Fig. 1b) \cite{Eva98}.
\begin{figure}
\centerline{
\psfig{figure=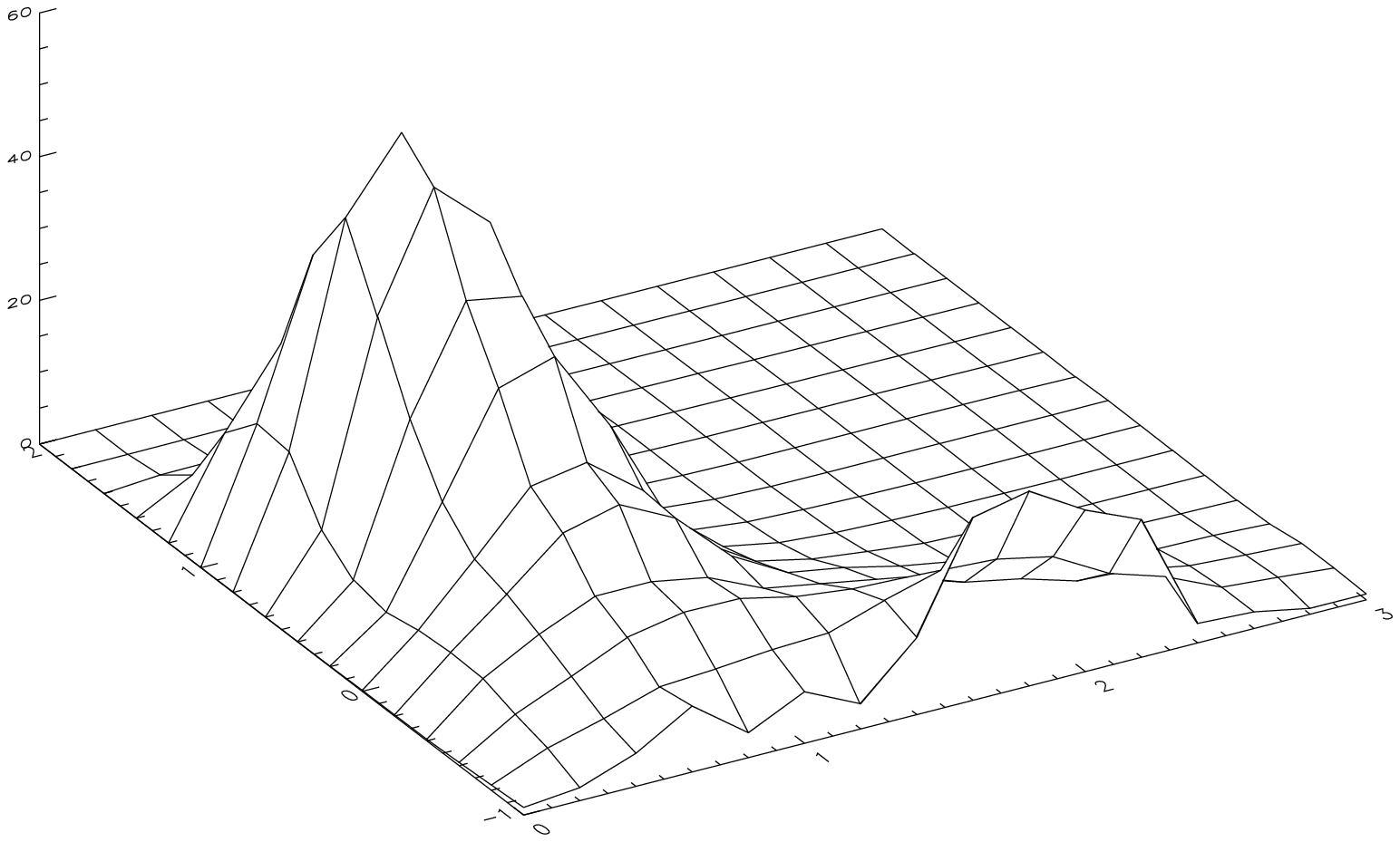,height=4cm,angle=0}
\psfig{figure=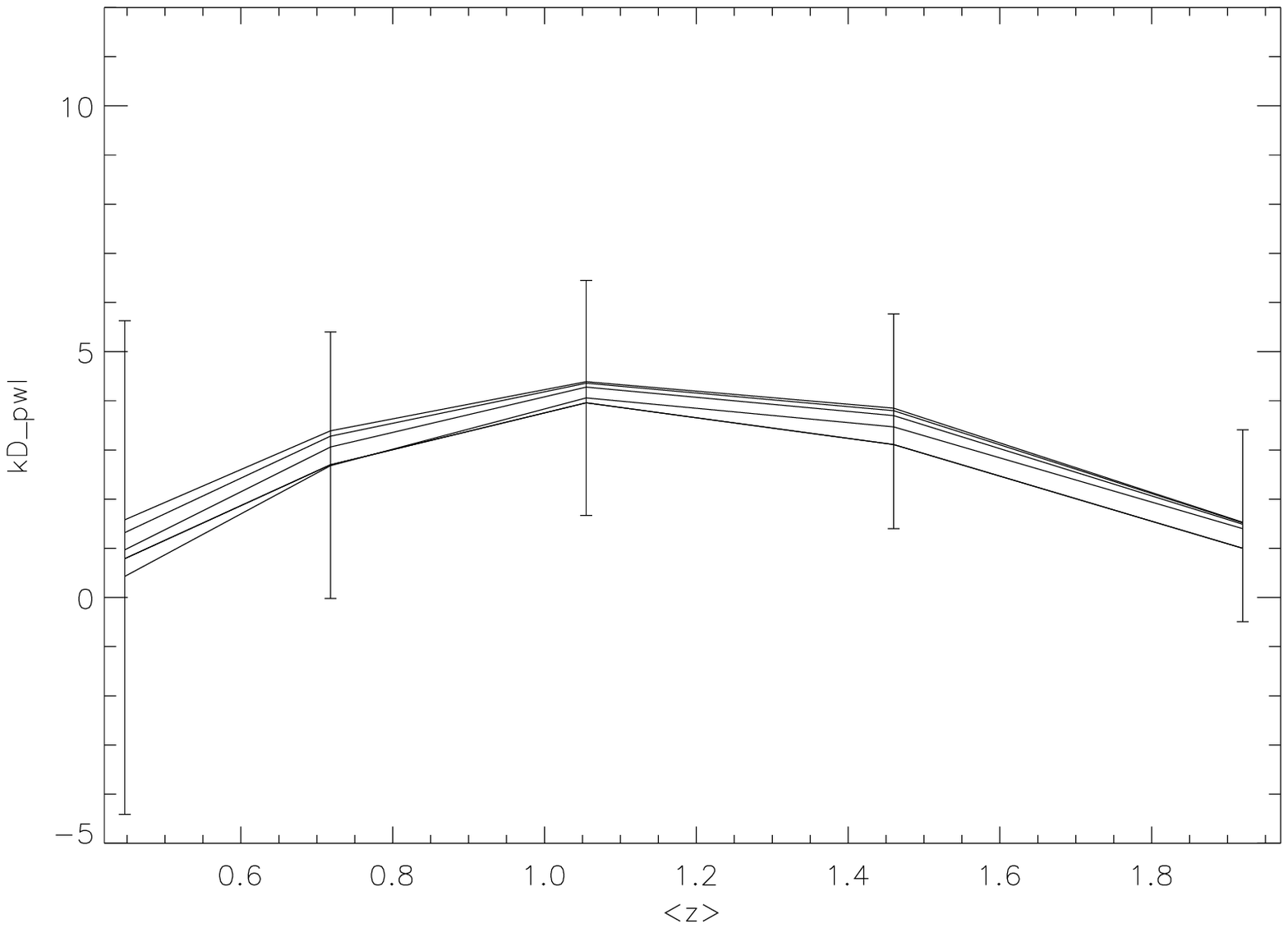,height=4cm,angle=0}
}
\caption{Left: Likelihood map in the $(\Omega,\Lambda)$ plane,
\cite{LVW96}. Right: PDE parameter in bins of redshift, \cite{Eva98}}
\end{figure}

So far, cosmological constraints from quasar samples have been 
obtained from simplistic evolution schemes only, 
however the actual evolution appears to be far more complex at
redshifts $2<z<3$. The theoretical progresses mentioned above will allow
better constraints to be extracted in this way from quasar samples complete 
at all redshifts.

\section{Multicolor selection, and the problem of 
the redshift range $2.5 \le z \le 3.2$ }
Before using quasar samples for cosmological purposes, one must beware
of observational biases.
Multicolor techniques such as UVX or BRX are very efficient in defining
quasar candidates in various redshift ranges. More precisely, the most
known UVX selection is extremely efficient for (blue) quasars up to
redshift $\sim 2.2$ (see for example \cite{B90}, \cite{Hall}), 
while BRX selection reveals quasar candidates at even higher redshift
($z \ge 3$, \cite{Hall}). However, in all recent attempts to
construct a complete catalog, a problem occurs for quasars whose
redshift belongs to the interval $2.5 \le z \le 3.2$ where the
multicolor techniques cannot usually distinguish between quasars and main
sequence stars, because of quasars' stellar-like colors. Using an
appropriate combination of filters this redshift interval can be
restrained but it has never been completely covered until now.

Fig. 2 shows the superposition of the (simulated) quasar trails and 
the (observed) main sequence stars \cite{Hall98} with $m_B \in 
[21,22.5]$ in the 
redshift ranges [2.5,2.8] and [3.5,3.7] in the -(B-V)/(V-R) plan.

\begin{figure}
\centerline{
\psfig{figure=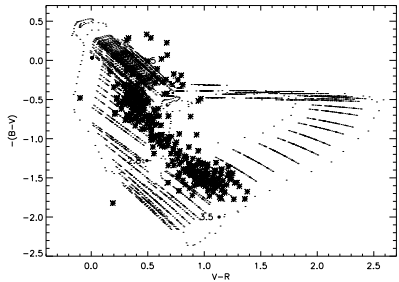,height=4cm,angle=0}}
\caption{-(B-V) versus (V-R) diagram: notice the superposition of the
(simulated) quasar trails and the (observed) main sequence stars with
$m_B \in [21,22.5]$ (data taken from \cite{Hall98})}
\end{figure}

\section{The VIRMOS quasars }
Several samples of typically a few thousand quasars, a size
quite convenient for most of the above purposes, 
are planned or are 
presently being assembled (see several contributions at this 
conference). Most of them, however, are based on the preselection of quasar
candidates, in order to optimize the efficiency of telescope time. 
Preselection makes the completeness questionable both at low redshift 
(morphological preselection of a stellar shape, 
incompleteness by a factor up to 5 
\cite{K}), and at high redshift (color 
preselection, results from color-color diagrams of \cite{Hall}
still reveals a strong bias in the range $2.2 \le z \le 3$, exactly
where a reversing of evolution is observed).

A promising observational strategy is allowed by the VIRMOS Survey,
which could avoid the main drawbacks of preselection.
To improve the efficiency of all quasar studies, one obviously
has to increase both the size and the upper redshift 
limit of the sample, and to be as
complete as allowed over the whole redshift range.
The VIRMOS spectroscopic large survey of {\it faint galaxies} 
will provide the opportunity of
assembling such a sample. This is because stars and 
stellar objects are only a marginal population of the deep sky, 
allowing the spectroscopy of a large fraction of stellar objects 
to be done.
At this conference, O. Le fevre already described the main characteristics
of this survey. For our concern, what is important with respect to most
previous or future quasar surveys, is that there will be NEITHER color
NOR morphological preselection of quasar candidates, but only some
random selection for the shallow sample. The obvious advantage will be
the possibility to test the completeness of the most current surveys, 
both at low and high redshift, which are based on morphological and color 
preselection, respectively. We expect to assemble 2 unbiased samples,
($1300^{+800}_{-500}$) quasars brighter than I=22.5 from the shallow survey
and ($700?^{+600?}_{-300?}$) quasars brighter than I=24 from the deep survey.
Furthermore we are examining the possibility to get a third, 
color-selected, sample of $4500^{+500}_{-500}$ quasars, 
which would be far larger, but biased because obtained by spectroscopy of 
{\it candidates} based both on color preselection in the associated
MEGAPRIME photometric survey and on counterparts of sources 
in the XMM Deep Survey (see the contribution of M. Pierre, this
conference)
which will have at least some common fields with MEGAPRIME
and VIRMOS surveys.

In brief, the VIRMOS quasar sample will offer the opportunity to check 
the usual modes of selection of quasar candidates, to study numerous
absorbers, to study jointly active galaxies and quasars in a complete
sample containing both with essentially common selection criteria,
and, last but not least, it will be partly free of biases in the
redshift range $2<z<3$ where evolution is specially complex.

\vfill
\end{document}